\documentclass[amsmath,amssymb,twocolumn,showpacs,superscriptaddress,prl,longbibliography]{revtex4-1}

\usepackage{dcolumn}% Align table columns on decimal point
\usepackage{bm}% bold math
\usepackage{graphicx}
\usepackage{epstopdf}
\usepackage{wrapfig}
\usepackage{hyperref}
\usepackage{array} 
\usepackage{listings}
\usepackage[para,online,flushleft]{threeparttablex}
\usepackage{booktabs,dcolumn}
\usepackage{filecontents}
\usepackage{breqn}
\usepackage{xcolor}
\usepackage{bibunits}
\usepackage{lipsum}
\usepackage{upgreek} %to have an upright greek symbol mu for microsecond
\usepackage{etoolbox}
\makeatletter
\newcommand*{\newbibstartnumber}[1]{%
  \apptocmd{\thebibliography}{%
    \global\c@NAT@ctr #1\relax
    \addtocounter{NAT@ctr}{-1}%
  }{}{}%
}
\makeatother
\makeatletter
\let\cat@comma@active\@empty
\makeatother

\begin{document}

\preprint{APS/123-QED}

\title{Isotope Shifts of Radium Monofluoride Molecules}% Force line breaks with \\

\author{S.M.~Udrescu} 
\email{sudrescu@mit.edu}
\affiliation{Massachusetts Institute of Technology,~Cambridge,
~MA~02139,~USA}
\author{A.J.~Brinson} 
\affiliation{Massachusetts Institute of Technology,~Cambridge,
~MA~02139,~USA}
\author{R.F. Garcia Ruiz}
\email{rgarciar@mit.edu}
\affiliation{Massachusetts Institute of Technology,~Cambridge,
~MA~02139,~USA}
\affiliation{CERN,~CH-1211~Geneva~23,~Switzerland}
\author{K. Gaul}
\affiliation{Fachbereich Chemie, Philipps-Universit{\"a}t~Marburg, Hans-Meerwein-Stra{\ss}e~4,~35032~Marburg,~Germany}
\author{R. Berger}
\email{robert.berger@uni-marburg.de}
\affiliation{Fachbereich Chemie, Philipps-Universit{\"a}t~Marburg, Hans-Meerwein-Stra{\ss}e~4,~35032~Marburg,~Germany}
\author{J.~Billowes}
\affiliation{School of Physics and Astronomy,~The~University~of~ Manchester,~Manchester~M13~9PL,~United~Kingdom}
\author{C.L.~Binnersley}
\affiliation{School of Physics and Astronomy,~The~University~of~ Manchester,~Manchester~M13~9PL,~United~Kingdom}
\author{M.L.~Bissell}
\affiliation{School of Physics and Astronomy,~The~University~of~ Manchester,~Manchester~M13~9PL,~United~Kingdom}
\author{A.A. Breier} 
\affiliation{Laboratory for Astrophysics,~Institute~of~Physics,~University~of~Kassel,~34132 ~Kassel,~Germany}
\author{K.~Chrysalidis}
\affiliation{CERN,~CH-1211~Geneva~23,~Switzerland}
\author{T.E.~Cocolios}
\affiliation{KU Leuven,~Instituut~voor~Kern-~en~Stralingsfysica, ~B-3001~Leuven,~Belgium}
\author{B.S.~Cooper}
\affiliation{School of Physics and Astronomy,~The~University~of~ Manchester,~Manchester~M13~9PL,~United~Kingdom}
\author{K.T.~Flanagan}
\affiliation{School of Physics and Astronomy,~The~University~of~ Manchester,~Manchester~M13~9PL,~United~Kingdom}
\affiliation{Photon Science Institute,~The~University~of~Manchester, ~Manchester~M13~9PY,~United~Kingdom}
\author{T.F. Giesen}
\affiliation{Laboratory for Astrophysics,~Institute~of~Physics,~University~of~Kassel,~34132 ~Kassel,~Germany}
\author{R.P.~de~Groote}
\affiliation{Department of Physics,~University~of~Jyv\"askyl\"a, ~Survontie~9,~Jyv\"askyl\"a,~FI-40014,~Finland}
\author{S. Franchoo}
\affiliation{Institut de Physique Nucleaire d'Orsay,~F-91406~Orsay, ~France}
\author{F.P.~Gustafsson}
\affiliation{KU Leuven,~Instituut~voor~Kern-~en~Stralingsfysica, ~B-3001~Leuven,~Belgium}
\author{T.A. Isaev}
\affiliation{NRC Kurchatov Institute-PNPI,~Gatchina,~Leningrad~ district~188300,~Russia}

\author{\'A.~Koszor\'us}
\affiliation{KU Leuven,~Instituut~voor~Kern-~en~Stralingsfysica, ~B-3001~Leuven,~Belgium}
\author{G.~Neyens}
\affiliation{CERN,~CH-1211~Geneva~23,~Switzerland}
\affiliation{KU Leuven,~Instituut~voor~Kern-~en~Stralingsfysica, ~B-3001~Leuven,~Belgium}

\author{H.A. Perrett}
\affiliation{School of Physics and Astronomy,~The~University~of~ Manchester,~Manchester~M13~9PL,~United~Kingdom}
\author{C.M.~Ricketts}
\affiliation{School of Physics and Astronomy,~The~University~of~ Manchester,~Manchester~M13~9PL,~United~Kingdom}
\author{S. Rothe}
\affiliation{CERN,~CH-1211~Geneva~23,~Switzerland}
\author{A.R.~Vernon}
\affiliation{School of Physics and Astronomy,~The~University~of~ Manchester,~Manchester~M13~9PL,~United~Kingdom}
\author{K.D.A.~Wendt}
\affiliation{Institut f\"ur Physik,~Johannes~Gutenberg-Universit\"at ~Mainz,~D-55128~Mainz,~Germany}
\author{F. Wienholtz}
\affiliation{CERN,~CH-1211~Geneva~23,~Switzerland}

\affiliation{Institut f\"ur Physik,~Universit\"at~Greifswald, ~D-17487~Greifswald,~Germany}
\author{S.G.~Wilkins}
\affiliation{CERN,~CH-1211~Geneva~23,~Switzerland}

\author{X.F.~Yang}
\affiliation{School of Physics and State Key Laboratory of Nuclear Physics and Technology,~Peking~University,~Beijing~100971,~China}

\date{\today}

\begin{abstract}
Isotope shifts of $^{223-226,228}$Ra$^{19}$F were measured for different vibrational levels in the electronic transition $A^{2}{}{\Pi}_{1/2}\leftarrow X^{2}{}{\Sigma}^{+}$. The observed isotope shifts demonstrate the particularly high sensitivity of radium monofluoride to nuclear size effects, offering a stringent test of models describing the electronic density  within the radium nucleus. \emph{Ab initio} quantum chemical calculations are in excellent agreement with experimental observations. These results highlight some of the unique opportunities that short-lived molecules could offer in nuclear structure and in fundamental symmetry studies. 
%Isotope shifts of $^{223-226,228}$Ra$^{19}$F were measured in the electronic transition $A^{2}{}{\Pi}_{1/2}\leftarrow X^{2}{}{\Sigma}^{+}$ with the Collinear Resonance Ionization Spectroscopy (CRIS) experiment at the ISOLDE facility at CERN. The observed isotope-dependent shifts of the vibronic transition energies demonstrate the particularly high sensitivity of radium monofluoride to nuclear size effects, offering a stringent test of models describing the electronic density  within the radium nucleus. \emph{Ab initio} quantum chemistry calculations are in excellent agreement with experimental observations. These results establish an important benchmark for the further development of quantum chemistry calculations, which are essential to extract nuclear properties and fundamental physics observables from high-precision molecular spectroscopy experiments. These findings highlight some of the unique opportunities that short-lived molecules containing elements from the seventh row of the periodic table could offer in nuclear structure and in fundamental symmetry studies. 

\end{abstract}

\maketitle

\textit{Introduction}. The study of the electron-nucleus interactions offers a powerful tool in the exploration of the atomic nucleus and of the fundamental particles and interactions of nature \cite{safronova2018search}. In recent years, the immense progress of theoretical and experimental molecular spectroscopy are breaking new ground in fundamental physics research. The structure of well-chosen molecular systems can offer exceptionally high sensitivity to investigate the violation of fundamental symmetries, which can be enhanced by more than five orders of magnitude with respect to atomic systems
\cite{flambaum2019enhanced, auerbach1996collective, kudashov2014ab, safronova2018search, PhysRevA.102.062801}.
Ongoing developments pave the way for distinct approaches to measure symmetry-violating effects that could rigorously test the Standard Model at low energy, and constrain the existence of new physics \cite{flambaum2014time,altuntacs2018demonstration,gaul2019systematic}.

Electronic states in atoms and molecules can be highly sensitive to the structure of their atomic nuclei, and enable detailed investigations into the electron-nucleon interaction \cite{dela17a,dela17b,berengut18,skripnikov2020nuclear}. Adding or removing neutrons to/from an atomic nucleus results in small differences in the energies of its bound electrons, known as isotope shifts. In atoms, isotope shift measurements have provided unique access to study the evolution of nuclear charge radii in exotic nuclei 
\cite{garciaruiz16,Campbell2016,Hag16,marsh2018characterization,groote20,kaufmann2020charge}. Precision measurements along isotopic chains can be used to separate electronic and nuclear effects, thereby placing powerful constraints on the violation of fundamental symmetries and the search for new physics beyond the Standard Model \cite{antypas2019isotopic,dela17a,dela17b,berengut18,counts2020evidence}. The possibility of performing precision measurements over long chains of isotopologues -- molecules of the same elements that differ by the number of neutrons in their nuclei -- offers an ideal scenario to investigate nuclear structure phenomena, nuclear-spin-dependent parity violation interactions, and probe fine details of the electron-nucleon interaction in so far unexplored regimes. 
%Remarkably, atomic parity violation measurements have established our best low-energy test of the Standard Model \cite{wood1997measurement,safronova2018search,antypas2019isotopic} . 

A separation of so-called mass and volume effects from isotope shift measurements requires at least two relative measurements (three isotopologues), but no heavy element beyond Pb possesses more than two long-lived isotopes \cite{sonzogni2005nudat,cocolios2017new}. This has been a major experimental obstacle,  limiting our knowledge of molecules containing heavy nuclei, as radioactive molecules can often only be produced in small quantities (typically less than $10^7$ molecules/s). Thus, their study requires exceptionally sensitive experimental techniques, which have not been available until very recently \cite{ruiz2019spectroscopy}.

%The present work reveals that the special electronic structure of radium monofluoride (RaF), which was theoretically predicted \cite{isaev2010laser} and experimentally confirmed \cite{ruiz2019spectroscopy} to be favourable for molecular cooling with lasers due to nearly diagonal Franck--Condon matrices \cite{dirosa2004,isaev2010laser,isaev2016poly,isaev2018review}, facilitates the determination of nuclear volume shifts even from mid- to low-resolution vibronic profiles. 

In parallel to experimental advances in molecular physics, the development of molecular theory is of critical importance. Accurate determination of molecular parameters is important not only to guide experiments, but also essential to extract nuclear structure and new physics observables from experimental results. The interpretation of molecular experiments relies upon computations of molecular enhancement parameters, which strongly depend on the description of the electron wave function at the nucleus. Hence, measurements of molecular isotope shifts, which are highly sensitive to the electron-nucleus interaction at the nucleus, provide an important test for the reliability of quantum chemical calculations. However, until now, comparatively little was known about electronic shifts in isotopologues \cite{mchale2017molecular}. To our knowledge, isotope shift measurements have neither been reported yet in molecules containing isotopes of elements heavier than lead (Pb, $Z=82$) \cite{kno1985molecular}, nor for molecules containing short-lived isotopes.

%This letter reports a first step in this direction. Measurements of transition energies of electronic transitions between different vibratonal states were performed for different isotopologues of radium monofluoride molecules, RaF. Isotope shifts measurements in these molecules allow exhibt high sensitivity to the nuclear charge radius of the heavy atom, offering compelling alternatives to investigate nuclear structure phenomena and probe theoretical descriptions of the electronic wave function at the nucleus. 

In this Letter we investigate, theoretically and experimentally, the changes in the molecular energy levels when neutrons are added to (or removed from) the radium nucleus of RaF molecules. RaF molecules are of special interest for fundamental physics research as their electronic structure is predicted to provide a large enhancement of both parity and time-reversal violating effects \cite{isaev2010laser, isaev2013lasercooled, kudashov2014ab, gaul2019systematic, gaul2020toolbox}. Moreover, some nuclei in the very long isotope chain of Ra ($Z=88$) exhibit octupole deformation \cite{gaffney2013studies,butler2020evolution}, which results in a significant amplification of their symmetry-violating nuclear properties, relative to light molecules \cite{flambaum2019enhanced,  auerbach1996collective}.

\begin{figure}[!htbp]
\includegraphics[width=\columnwidth,height=0.725\textheight]{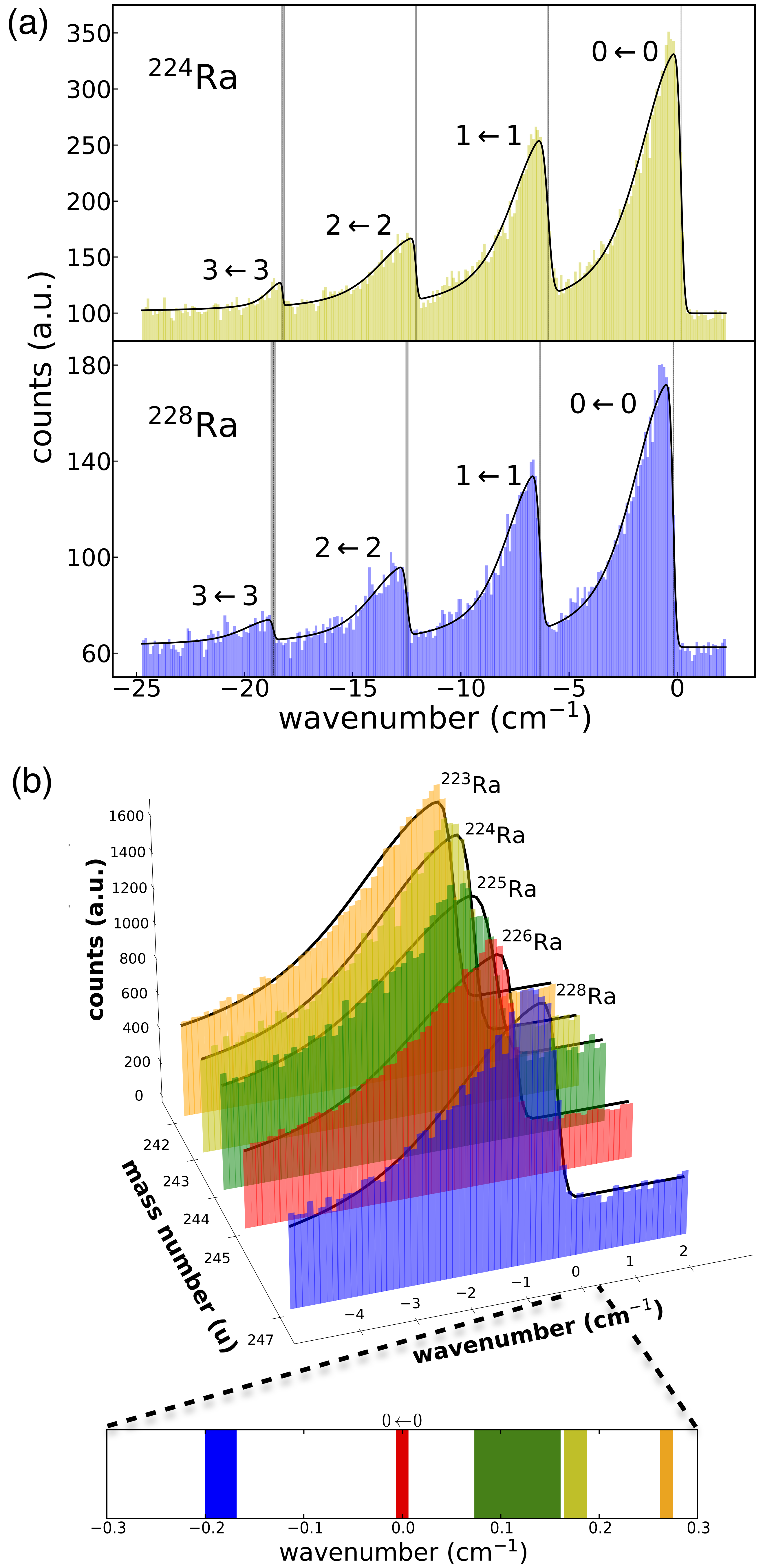}
\caption{\label{spectra_fit} Iso-vibrational spectra of the $A^2{\Pi}_{1/2}\leftarrow X^2{\Sigma}^{+}$ electronic transition. (a) Spectra  of the $^{224}$Ra$^{19}$F (top) and $^{228}$Ra$^{19}$F (bottom) isotopologues. The wavenumber axis was shifted such that the origin coincides with the center of the $0 \leftarrow 0$ transition of the reference $^{226}$Ra$^{19}$F isotopologue.  Moving in the negative direction on the wavenumber axis, the four peaks correspond to the $0 \leftarrow 0$, $1 \leftarrow 1$, $2 \leftarrow 2$ and $3 \leftarrow 3$ vibrational transitions. The histograms
represent the experimental data, while the continuous black
curves depict the best fits for each isotopologue. The vertical bands mark the central values of the transitions, while the width of each band corresponds to the associated uncertainty. (b) $0 \leftarrow 0$ peaks of the five isotopologues $^{223-226,228}$Ra$^{19}$F. The plot at the bottom marks the position of the center of each peak relative to the one corresponding to $^{226}$Ra$^{19}$F. The width of each vertical band shows the uncertainty on the given central value.}
\end{figure}

\textit{Experimental Technique}. Our experimental approach was described in detail in Ref.~\cite{ruiz2019spectroscopy}. Briefly, Ra isotopes were produced at the ISOLDE facility at CERN, by impinging 1.4-GeV protons onto a uranium-carbide target. Upon injection of CF$_4$ gas into the target material, radium molecules were formed through reactive collisions of the evaporated Ra atoms. RaF$^{+}$ molecules were created by surface ionization and extracted using an electrostatic field. The desired isotopologue was selected using a high-resolution magnetic mass separator (HRS). After that, ions were collisionally cooled for up to 10 ms in a radio-frequency quadrupole (RFQ) trap, filled with helium gas at room temperature. Bunches of RaF$^{+}$ of 4~$\upmu$s temporal width were released and accelerated to 39998(1) eV, and then sent to the Collinear Resonance Ionization Spectroscopy (CRIS) setup \cite{flanagan2013collinear, de2015use, ruiz2018high, vernon2020optimising, agota2019resonance}.

At the CRIS beam line, the ions were passed through a charge-exchange cell where they were neutralized in-flight by collisions with a sodium vapor, primarily by the reaction RaF$^{+}$+Na$\to$ RaF + Na$^{+}$. The estimated ionization energy of RaF is close to that of the atomic Na ($5.14$ eV), hence the RaF molecules predominately populated the $X^{2}\Sigma^+$ electronic ground state in the neutralization process \cite{isaev2013ion}. After this step, any remaining ions were deflected from the main, neutral molecular beam, which was then collinearly overlapped in space and time with two pulsed laser beams in an ultra-high-vacuum ($10^{-10}$ mbar) laser-molecule interaction region.

\begin{figure}[t]
\includegraphics[scale=0.165]{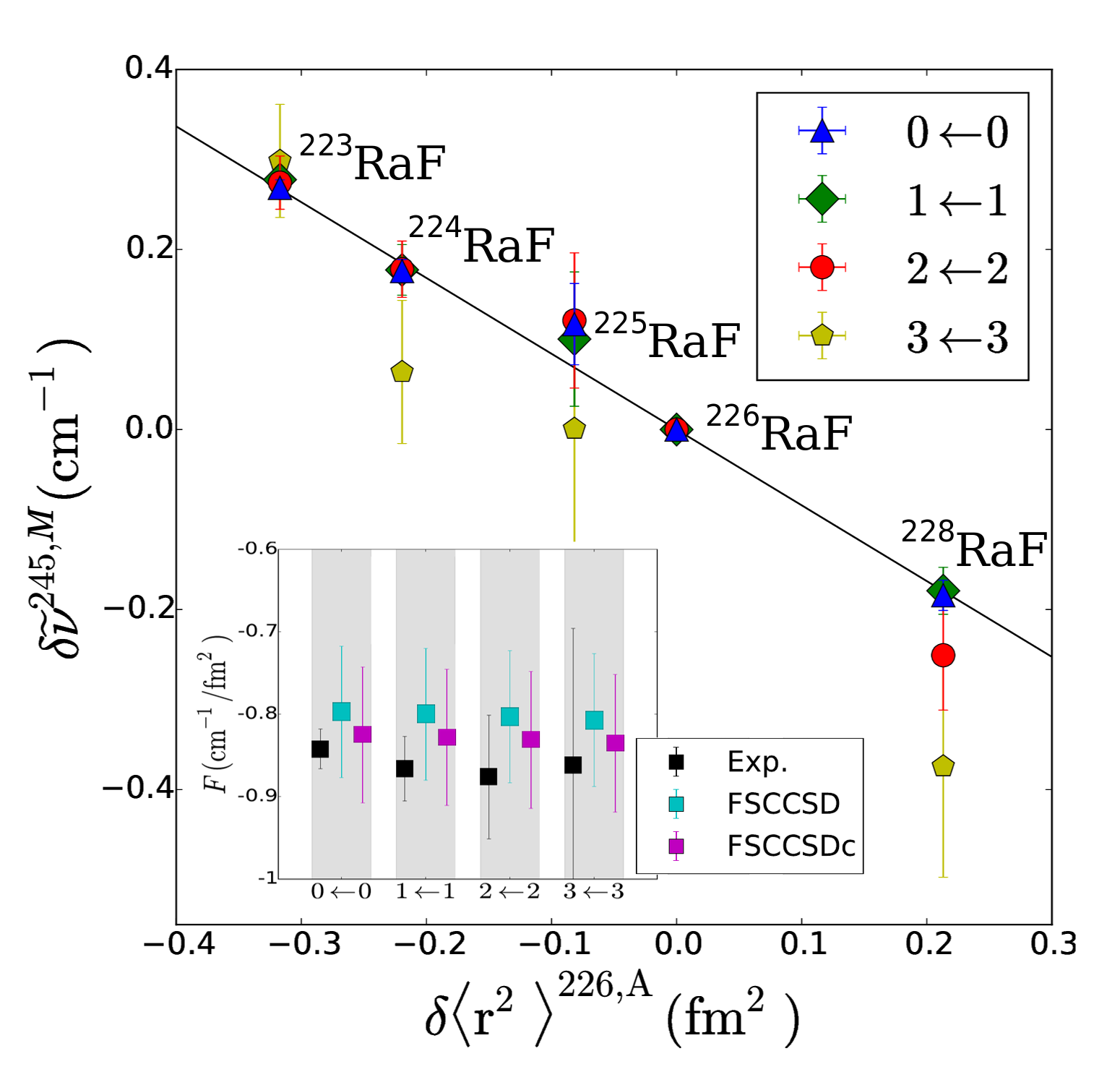}
\vspace{-0.8cm}
\caption{\label{is_shift_vs_rad}  Isotope shifts as a function of the changes in the charge radius of the five analyzed RaF isotopologues with respect to the reference $^{226}${Ra$^{19}$F} molecule. Each color corresponds to one of the four investigated transitions and the black line shows the best linear fit to the $0 \leftarrow 0$ transition. The inset in the bottom left shows the values of the slope of this fit, $F$, corresponding to experimental data (black) and theoretical predictions (cyan for FSCCSD method and magenta for FSCCSDc). See main text for details.}
\end{figure}

The RaF molecules were resonantly ionized in two steps. Firstly, one laser pulse of tunable wavelength was used to resonantly excite the electronic transition of interest, $A^{2}{}{\Pi}_{1/2} \leftarrow X^{2}{}{\Sigma}^{+}$. The electronically excited RaF molecules were then ionized using a high-power, 355-nm pulsed laser. Subsequently, the resonantly ionized RaF$^+$ ions were deflected from the bunch and detected by an ion detector. The wavelength of the ionization laser was set such that RaF molecules can be ionized if they are already in an excited electronic state, i.e. if the frequency of the first laser was on resonance with a transition in the RaF molecule. Hence, the low-lying electronic and vibrational spectra were obtained by counting the number of ions detected as a function of the wavelength of the first laser pulse.  

%More details on the experimental technique can be found in Ref. \cite{ruiz2019spectroscopy}.

\textit{Computational Methods}. Electronic transition wavenumbers $\tilde{\nu}$ for different isotopologues of RaF were calculated
at the level of relativistic Fock-Space Coupled Cluster including Singles and Doubles
amplitudes (FSCCSD) with the program
package DIRAC19~\cite{DIRAC19}, correlating 17 electrons (FSCCSD(17e)). Isotope shift constants $F=\frac{\partial\delta\tilde{\nu}}{\partial\delta\langle{r^2}\rangle}$ were deduced by calculation of transition wavenumbers for different mean-square nuclear charge radii $\langle{r^2}\rangle$ with a Gaussian nuclear charge distribution model and computation of the slope within a linear fit model. 

In order to estimate the quality of this method, FSCCSD calculations
with an extended basis set and 27
correlated electrons (FSCCSD(27e)) were
carried out for three nuclear charge radii at an internuclear distance of 4.3$a_0$ (close to the RaF ground state bond length, see Ref.~\cite{isaev2013lasercooled}). This extended basis set was also used in atomic calculations of Ra$^+$, in which 19 electrons (5d, 6s, 6p and 7s shells) were correlated. These atomic calculations were used to directly compare our measurements and previous isotope shift measurements in the radium ion \cite{wansbeek:2012charge}. To account for the effect of the larger basis and active space in the other calculations, the FSCCSD(17e) isotope shifts were corrected by a factor $
\frac{F_{
\mathrm{FSCCSD(27e)}}(4.3\,a_0)}{F_{
\mathrm{FSCCSD(17e)}}(4.3\,a_0)}$, to which we will refer as FSCCSDc in the following.

Isotope shift constants $F(r_{\mathrm{RaF}})=\frac{\partial\delta \widetilde{\nu}}{\partial\delta\langle r^2
\rangle}(r_{\mathrm{RaF}})$ were calculated for bond lengths of $r_{\mathrm{RaF}}$ = 4.0,  4.1,  4.2,  4.25,  4.3 and  4.4~$a_0$, which covers the region around the equilibrium structure of the electronic ground state. From this, we determined $F$ as a function of $r_{\mathrm{RaF}}$ from a fourth-degree polynomial fit. Vibrational corrections to $F$ were calculated within a one-dimensional discrete variable representation (DVR) approach. In DVR calculations, potentials of the electronic ground and excited states calculated in Ref.~\cite{isaev2013lasercooled} were employed. For more details on the applied active space in FSCCSD calculations, used basis sets, the employed Gaussian model and the vibrational corrections see the Supplemental Material \cite{supplmat}.

\textit{Results and Discussion}. The measured iso-vibrational spectra of the $A^{2}{}{\Pi}_{1/2}\leftarrow X^{2}{}{\Sigma}^{+}$ electronic manifold for two of the investigated isotopologues, $^{224}$Ra$^{19}$F (top) and $^{228}$Ra$^{19}$F (bottom), are shown in Fig. \ref{spectra_fit}(a). The reported wavenumbers have been transformed to the molecular rest frame to account for the Doppler shift between the lab frame and the molecular bunch velocity. Each spectrum was fitted with a sum of four skewed-Voigt profiles in addition to a constant background, from which the centers of the $0 \leftarrow 0$, $1 \leftarrow 1$, $2 \leftarrow 2$ and $3 \leftarrow 3$ transitions were extracted. Wavenumber values are shown relative to the $0 \leftarrow 0$ transition of $^{226}${Ra$^{19}$F}. The continuous curves show the best fits to the data. The vertical lines indicate the central value of each of the four transitions, with the widths indicating the associated uncertainty. A close-up view of the spectra corresponding to the $0 \leftarrow 0$ transition of the five isotopologues $^{223-226,228}$Ra$^{19}$F can be seen in Fig. \ref{spectra_fit}(b). The plot at the bottom shows the position of the center of each peak relative to the one corresponding to $^{226}$Ra$^{19}$F. Each vertical band corresponds to a given isotopologue according to the color used in the top plot, and the width of the band shows the uncertainty on the given central value. A shift of the central values between the measured isotopologues can be clearly observed. The obtained isotope shifts  for all five isotopologues are shown in Table \ref{isotope_shifts_MHz}.

\begin{table}
\caption{\label{isotope_shifts_MHz}
Measured isotope shifts of the $A^{2}{}{\Pi}_{1/2}\leftarrow X^{2}{}{\Sigma}^{+}$ vibrational  transitions for the isotopologues $^{223-226,228}${}{Ra$^{19}$F} (in units of $\mathrm{cm^{-1}}$). The isotope shifts are given relative to $^{226}$Ra$^{19}$F.
}
\begin{ruledtabular}
\begin{tabular}{ccccc}
%\multicolumn{1}{c}{\textrm{$A$}}&
\textrm{\textrm{$M$}}&
\textrm{\textrm{$\delta\widetilde{\nu}^{245,M}_{0 \leftarrow 0} $}}&
\textrm{\textrm{$\delta\widetilde{\nu}^{245,M}_{1 \leftarrow 1}$}}&
\textrm{\textrm{$\delta\widetilde{\nu}^{245,M}_{2 \leftarrow 2}$}}&
\textrm{\textrm{$\delta\widetilde{\nu}^{245,M}_{3 \leftarrow 3}$}}\\
\colrule
242 & $0.269(9)$   & $0.278(14)$   & $0.274(30)$   & $0.298(63)$\\
243 & $0.176(13)$   & $0.177(28)$  & $0.178(31)$   & $0.064(80)$\\
244 & $0.117(46)$  & $0.100(74)$  & $0.121(75)$   & $0.001(131)$\\
245 & $0$          & $0$          & $0$           & $0$\\
247 & $-0.184(17)$ & $-0.179(26)$ & $-0.251(60)$  & $-0.374(123)$\\
\end{tabular}
\end{ruledtabular}
\end{table}

%, together with the associated errors. The errors come from the resulting parameter uncertainties estimated by the fitting procedure, and from taking into account the variations in the estimated centers due to the frequency resolution used in binning the raw data. 

At a bond length of $4.3~a_0$, the isotope shift constant $F$ is computed to be $-0.797$ at the level of FSCCSD(17e) and $-0.825$ at the level of FSCCSD(27e), respectively. From a comparison of these two values we estimate the relative uncertainty of the FSCCSD(17e) approach due to the basis set and the size of the active space to be less than $4~\%$. Comparison of the isotope shift constant in the $7p^2P_{1/2}\leftarrow 7s^2S_{1/2}$ transition in Ra$^+$ calculated at the level of FSCCSD(27e) ($F=-1.283(3)\,\frac{\mathrm{cm}^{-1}}{\mathrm{fm}^2}$) to atomic calculations with a comparable method but a different nuclear model in Ref.~\cite{wansbeek:2012charge} ($F=-1.328\,\frac{\mathrm{cm}^{-1}}{\mathrm{fm}^2}$) shows a deviation of about $3~\%$. Assuming that this deviation is mainly due the use of a different model for the nuclear charge distribution we estimate the error due to the Gaussian nuclear model to be about $3~\%$. Thereby, we note that the shape of the potential of a Gaussian nuclear charge distribution is rather similar to the shape of the potential of a homogeneously charged solid sphere nuclear model used as starting point in Ref.~\cite{wansbeek:2012charge}, as well as to the shape of the potential of a Fermi nuclear model (see Ref.~\cite{andrae2000finite} for a review). Finally, from CC calculations of hyperfine couplings in the $X^{2}{}{\Sigma}^{+}$ state \cite{kudashov2014ab} and the $A^{2}{}{\Pi}_{1/2}$ state \cite{skripnikov2020nuclear} we assume contributions of excitations involving three electrons as characterized by the $T_3$ CC amplitudes to be less than $3~\%$. Errors from other sources, such as the vibrational description employed in comparison are expected to be negligible. Thus, an overall uncertainty of about 10~\% is estimated for the calculated isotope shift constants. 

Results of \emph{ab initio} calculations of isotope shift constants $F$ for the electronic states $X^2\Sigma_{1/2}$ and $A^2\Pi_{1/2}$ for the first four vibrational levels in RaF and a comparison to experimentally determined isotope shift constants (see details below) are shown in Table~\ref{theory_data}. Within the 10~\% uncertainty of the theoretical methods discussed above, the predictions are in agreement with experiment. This is also shown in the inset of Fig. \ref{is_shift_vs_rad}.

\begin{table}
\caption{\label{theory_data}
Calculated isotope shift constants $F=\frac{\partial\delta\tilde{\nu}}{\partial\delta\langle{r^2}\rangle}$ of vibrational levels $v$ for the $X^2\Sigma_{1/2}$ and $A^2\Pi_{1/2}$ electronic states are shown at the FSCCSD(17e) level of theory, with vibrational corrections at the level of DVR. FSCCSDc values shown in squared brackets in the fourth column. Comparison of the theoretically determined isotope shift constants to experimentally obtained $F$ are presented in the last two columns. Theoretical values are estimated to have a relative error of approximately $10\%$ (see text).}
\begin{ruledtabular}
\begin{tabular}{lllll}
$v$&
\multicolumn{3}{c}{$F~\left(\frac{\mathrm{cm^{-1}}}{\mathrm{fm}^2}\right)$}
\\
&$X^2\Sigma_{1/2}$&$A^2\Pi_{1/2}$&
\multicolumn{2}{c}{$A^2\Pi_{1/2},v\leftarrow X^2\Sigma_{1/2},v$}\\
&&&theor.&exp.\\
\colrule
0  &  $0.764$ & $-0.033$ & $-0.797$ [$-0.825$] & $-0.845(24)$ \\
1  &  $0.766$ & $-0.034$ & $-0.800$ [$-0.828$] & $-0.868(39)$\\
2  &  $0.769$ & $-0.035$ & $-0.803$ [$-0.831$] & $-0.876(75)$\\
3  &  $0.771$ & $-0.036$ & $-0.807$ [$-0.835$] & $-0.862(166)$\\
\end{tabular}
\end{ruledtabular}
\end{table}

Similar to atoms, the isotopic shifts of the electronic energy levels in molecules can be highly sensitive to the electron density at the nucleus and to changes of the nuclear size \cite{schlembach1982isotopic,knecht2012nuclear,almoukhalalati2016nuclear}. This in turn can provide valuable constraints for quantum chemical calculations, such as determination of the ground-state electronic wavefunction. However, molecules also possess vibrational and rotational degrees of freedom \cite{banwell1994fundamentals,barrow1975molecular,mchale2017molecular}. Changes in the nuclear volume between different isotopes can yield measurable deviations to the parameters associated with these degrees of freedom. Vibrational and rotational isotope shifts are sensitive to the first and second derivatives of the electronic density with respect to internuclear distance \cite{schlembach1982isotopic,knecht2012nuclear,almoukhalalati2016nuclear}.

By accounting for the finite nuclear size in addition to the breakdown of the Born-Oppenheimer approximation, the isotope shift can be approximately related to changes in the nuclear charge radius (see Supplementary Material \cite{supplmat} for derivation and further details) \cite{ross1974heterodyne,bunker1977nuclear,watson1980isotope,schlembach1982isotopic,le1999improved,knecht2012nuclear,almoukhalalati2016nuclear}:

% \begin{equation} 
%     \delta \nu^{\alpha,\alpha',\Pi,\Sigma,\nu} = \left(\Delta V_{00}^{\alpha,\Pi-\Sigma} +  \frac{\Delta V_{10}^{\alpha,\Pi-\Sigma}}{\sqrt{\mu_\alpha}}(\nu+1/2)\right)
%     \delta\langle r^2\rangle_{\alpha\alpha'} 
%     \label{eq1}
% \end{equation}

\begin{equation} 
    \delta \nu^{A,A',\Pi,\Sigma,\nu} = \left(\Delta V_{00}^{A,\Pi-\Sigma} +  \frac{\Delta V_{10}^{A,\Pi-\Sigma}}{\sqrt{\mu_A}}(\nu+1/2)\right)
    \delta\langle r^2\rangle_{AA'} 
    \label{eq1}
\end{equation}
where $\mu_A$ is the reduced mass of the reference isotopologue, $\Delta V_{00}^A$ and $\Delta V_{10}^A$ represent corrections to the $Y_{00}$ and $Y_{10}$ Dunham parameters due to finite nuclear size. These are related to the effective electronic density at the Ra nucleus, $\bar{\rho}_e^A$, as well as the first and second derivatives of this density with respect to the internuclear distance (see Ref. \cite{almoukhalalati2016nuclear}, for a detailed discussion).

%data reported in Ref. \cite{lynch2018laser} for the values of the isotope shifts of the radium isotopes, $\delta \nu^{226,A}_{714}$,together with their calculated field ($F_{714}$) and mass ($M_{714}$) shift factors, the changes relative to $\prescript{226}{}{Ra}$ in the mean-square charge radii, $\delta \langle r^2 \rangle^{226,A}$,  were calculated using:

%\begin{center}
%\begin{equation}
%\delta \nu^{226,A}_{714} = F_{714}\delta \langle r^2 \rangle^{226,A} + M_{714}\frac{A-226}{226A}
%\end{equation}
%\end{center}

Empirical values of the molecular parameters in Eq. \ref{eq1} can be extracted from the dependence of the measured molecular isotope shifts on the change in the mean-square charge radius, $\delta\langle r^2\rangle$, of the Ra nucleus \cite{wansbeek:2012charge,lynch2018laser}. The values used for $\delta\langle r^2\rangle$ were obtained from literature values of the $7p^2P_{1/2}\leftarrow 7s^2S_{1/2}$ transition measured in Ra$^+$~\cite{wansbeek:2012charge}, and using our calculated field shift, $F=-1.283(3)\,\frac{\mathrm{cm}^{-1}}{\mathrm{fm}^2}$.
%Using this value and the measured isotope shifts for the Ra$^+$ for the same transition from , we computed the changes in the charge radius $\delta\langle r^2\rangle $ between $^{226}$Ra and the other four isotopes considered in this analysis, using $\delta\tilde{\nu}^{226,A} = F\delta\langle r^2\rangle^{226,A}$. 
Mass-shift contributions to the isotope shift which are below 30 MHz were neglected and were included as an additional uncertainty. Fig. \ref{is_shift_vs_rad} shows the four measured transitions for RaF isotopologues as a function of the changes in the root-mean-square charge radius, $\delta \langle r^2 \rangle^{226,A}$, of the Ra nucleus with respect to the $^{226}$Ra isotope.

A fit to Eq. \ref{eq1} was performed for each of the four transitions. The extracted molecular parameters and theoretical results are shown in Table \ref{is_parameters}. The change in electron density at the nucleus between the $\Pi$ and $\Sigma$ states was obtained using $V_{00}^A =\frac{Z_Ae^2}{6\epsilon_0}\bar{\rho}_e^A $, 
 with $Z_A$ being the atomic number of the Ra nucleus, $e$ the electron charge, and $\epsilon_0$ the electric constant \cite{knecht2012nuclear,almoukhalalati2016nuclear}.
As seen in Table \ref{is_parameters}, an excellent agreement between experimental and theoretical results is obtained.  The isotope shifts in these molecules are dominated by the product between the changes of the nuclear size, $\delta \langle r^2 \rangle$, and the molecular parameter  $\Delta(V_{00}^A)^{\Pi-\Sigma}=-0.839(33)$ cm$^{-1}$/fm$^2$. This notably large sensitivity to nuclear size effects indicates that even for relatively low precision isotope shift measurements (uncertainty better than $0.02$ cm$^{-1}$), values of $\delta \langle r^2 \rangle$ could be extracted with better than 10 \% precision. Thus, such kind of molecules combined with the present method could offer sensitive routes to investigate nuclear structure properties of exotic actinide nuclei. These elements are highly reactive and challenging to produce and study in their atomic or ionic forms. However, their chemical properties facilitate the formation of molecular compounds. Hence, the extension of these studies to actinide molecules could provide new access to yet-unexplored nuclear properties in these nuclei.

\begin{table}
\caption{\label{is_parameters}Corrections to the $Y_{00}$ and $Y_{10}$ Dunham parameters due to the finite size of the nucleus (in units of $\mathrm{cm^{-1}/fm^2}$, and $\mathrm{u.^{1/2}cm^{-1}/fm^2}$, respectively). The last column shows the variation of the effective electron density between the $\Pi$ and $\Sigma$ states (in units of $\mathrm{\AA^{-3}}$).}
\begin{ruledtabular}
\begin{tabular}{cccc}
%\multicolumn{1}{c}{\textrm{$A$}}&
Method&
\textrm{\textrm{$\Delta V_{00}^{A,\Pi-\Sigma}$}}&
\textrm{\textrm{$\Delta V_{10}^{A,\Pi-\Sigma}$}}&
\textrm{\textrm{$\Delta(\bar{\rho}_e^A )^{\Pi-\Sigma}$}}\\
\colrule
exp. & -0.839(33) & -0.065(120)  & 392(15) \\
FSCCSD & -0.795(82) & -0.014(150)  & 371(38) \\
FSCCSDc & -0.823(85) & -0.014(155)  & 385(40) \\
\end{tabular}
\end{ruledtabular}
\end{table}

\textit{Summary and Outlook}. Isotope shift measurements of different RaF isotopologues were determined for the first time, exhibiting a remarkably high sensitivity to changes in the nuclear charge radius of the Ra nucleus. The effective electronic density at the Ra nucleus was obtained by combining our results with previous independent measurements of the mean-square charge radii of Ra isotopes. \emph{Ab initio} quantum chemical calculations show an excellent agreement with the experimental findings, confirming the high sensitivity that this molecule offers to study the electron-nucleon interaction in the proximity of the Ra nucleus.

%Future higher resolution measurements will allow the investigation of the rotational isotope shift of the RaF isotopologues. This can be used to constrain the values of the the first and second derivatives of the effective electronic density at the Ra nucleus with respect to the internuclear distance, which will further improve the theoretical calculations and allow for new physics searches using heavy molecules. For example, given the big number of rotational transitions available, searches for low mass bosons mediating a possible fifth fundamental force between the electrons and neutrons could be envisioned, by looking for non-linearities in a King-like plot \cite{frugiuele2017constraining,berengut2018probing,flambaum2018isotope,counts2020observation}.

 As illustrated in this work, measurements of molecular isotope shifts offer complementary information to test the reliability  of quantum chemical calculations, and provide empirical input by which the accuracy of theoretical models can be gauged. Quantum chemical calculations are essential to extract nuclear structure and fundamental physics parameters from precision measurements. Such information is especially interesting for exotic actinide species. These nuclei and their molecular compounds are of marked interest for nuclear structure and fundamental symmetry studies \cite{flambaum1997time,flambaum2008electric,engel2013electric,baron2014order,dobaczewski2018correlating,andreev2018improved,flambaum2019enhanced, skripnikov2020actinide, flambaum2020electric,Ga21,Fan21,Yu21, skripnikov2020nuclear}, but until now the experimental knowledge of their molecular and nuclear properties is lacking.  Our findings offer important information to understand nuclear structure and symmetry-violating effects in molecules containing the heaviest elements of the periodic table.

 %Until now, the experimental knowledge of. These elements are highly reactive and particularly challenging to produce in atomic or ionic forms. Hence, future laser spectroscopy studies of actinide molecules will offer new possibilities for the study of their nuclear properties. Precision measurements of short-lived radioactive molecules such as ThO will be of great interest for nuclear structure and studies fundamental symmetries   

\begin{acknowledgments}

This work was supported the ERC Consolidator Grant No. 648381 (FNPMLS); the Office of Nuclear Physics, U.S. Department of Energy, under grants DE-SC0021176 and DE-SC0021179; the MISTI Global Seed Funds; 
Deutsche Forschungsgemeinschaft (DFG, German Research Foundation) -- Projektnummer 328961117 -- SFB 1319 ELCH;
STFC grants ST/L005794/1, ST/P004423/1 and ST/L005786/1 and Ernest Rutherford Grant No. ST/L002868/1; projects from FWO-Vlaanderen, GOA 15/010 from KU Leuven and BriX IAP Research Program No. P7/12; the European Unions Grant Agreement 654002 (ENSAR2); the Russian Science Foundation under grant N 18-12-00227 (year 2020) and 21-42-04411 (year 2021); the BMBF grants 05P15HGCIA and 05P18HGCIA. The National Key RD Program of China (No: 2018YFA0404403) and the National Natural Science Foundation of China (No:11875073). We thank J. P. Ramos, J. Ballof and T. Stora for their support in the production of RaF molecules. We would also like to thank the ISOLDE technical group for their support and assistance. 
Computer time provided by the Center for Scientific Computing (CSC) Frankfurt is gratefully acknowledged.
\end{acknowledgments}

\bibliography{apssamp}% Produces the bibliography via BibTeX.

\end{document}